\documentclass[prd,aps,showpacs,nofootinbib,eqsecnum,onecolumn,groupedaddress,amssymb]{revtex4}
\usepackage{amsmath}
\usepackage{amssymb}
\usepackage{amsfonts}
\usepackage{graphicx,bm}
\usepackage{dcolumn}
\usepackage{color,amsxtra}
\usepackage{epsf}
\usepackage{enumerate}
\usepackage{hhline}
\usepackage{array}
\usepackage{tabularx}
\usepackage{subfigure}
\usepackage{fancyhdr}
\usepackage{mathrsfs}

\newcommand{\be}{\begin{equation}}
\newcommand{\ee}{\end{equation}}
\newcommand{\bea}{\begin{eqnarray}}
\newcommand{\eea}{\end{eqnarray}}
\newcommand{\beaa}{\begin{eqnarray*}}
\newcommand{\eeaa}{\end{eqnarray*}}

\def\be{\begin{equation}}
\def\ee{\end{equation}}
\def\bea{\begin{eqnarray}}
\def\eea{\end{eqnarray}}

\begin{document}

\title{Inflationary universe from higher derivative quantum gravity
coupled with scalar electrodynamics}
\author{
R.~Myrzakulov$^{1}$,
S.~D.~Odintsov$^{2, 3, 4}$
and
L.~Sebastiani$^{1}$
}
\affiliation{
$^1$Department of General \& Theoretical Physics and Eurasian Center for
Theoretical Physics, Eurasian National University, Astana 010008,
Kazakhstan\\
$^2$Consejo Superior de Investigaciones Cient\'{\i}ficas, ICE/CSIC-IEEC,
Campus UAB, Facultat de Ci\`{e}ncies, Torre C5-Parell-2a pl, E-08193
Bellaterra (Barcelona), Spain\\
$^3$ Institut de Ciencies de l'Espai (IEEC-CSIC),
Campus UAB, Carrer de Can Magrans, s/n
08193 Cerdanyola del Valles, Barcelona, Spain\\
$^4$ Tomsk State Pedagogical University, 634050 Tomsk, Russia\\ }

\begin{abstract}
We study inflation for a quantum scalar
electrodynamics  model in curved space-time and for higher-derivative quantum
gravity (QG) coupled with scalar electrodynamics.
The corresponding renormalization-group (RG) improved potential is
evaluated for both theories in Jordan frame where non-minimal
scalar-gravitational coupling sector is explicitly kept. The role of
one-loop quantum corrections is investigated by showing how these
corrections enter in the expressions for the slow-roll parameters, the
spectral index and the tensor-to-scalar ratio and how they influence the
bound of the Hubble parameter at the beginning of the primordial
acceleration. We demonstrate that the viable inflation maybe
successfully realized, so that it turns out to be consistent with last
Planck and BICEP2/Keck Array data.
\end{abstract}

\hspace{13.1cm}

\maketitle

\def\thesection{\Roman{section}}
\def\theequation{\Roman{section}.\arabic{equation}}

\section{Introduction}

Recent corrected Planck data as well as latest BICEP2/Keck/Array data
propose better quantative description of the inflationary universe. In its
own turn, this increases the interest to theoretical models of inflation
(for the reviews, see Ref.~\cite{inflrev}) because they maybe better
confronted against observational data.

During last years, there were many attempts to take into account  quantum
effects in order to construct
viable inflation in perturbative Einstein QG (for some review, see Ref.~\cite{Wo}). It is quite natural to go beyond semi-classical General Relativity
and to investigate  the
inflationary scenario for multiplicatively-renormalizable higher
derivative gravity as well as for string-inspired gravities.
The explicit calculation in this direction at strong gravity regime of
higher-derivative QG was done in Ref.~\cite{nostroHG} where possibility of
viable QG-induced inflation was proved. Of course, being the
multiplicatively-renormalizable
theory what gives the chance to evaluate QG corrections, higher-derivative
QG  represents merely the effective theory. It is known, that in such
theory the unitarity problem which is related with the
Ostrogradski instability~\cite{Wo1} remains to be the open issue.
Eventually,  in  higher-derivative gravity the
unitarity maybe restored at the non-perturbative level.  Thus,  this
theory could be considered as good approximation for the effective theory of
quantum gravity. One can expect to account for QG effects at least
qualitatively within such theory.

The purpose of this work is to study higher-derivative QG effects for
Higgs-like inflation. As simplified model we take first massless scalar
electrodynamics and investigate RG-improved inflation in such theory.
At the next stage, we consider higher-derivative QG coupled to scalar
electrodynamics and evaluate the corresponding RG-improved effective
potential. The ocurrence of viable inflation which is realized thanks to
such RG-improved effective potential with account of QG effects is proved.

The paper is organized in the following way. In Section {\bf II} we
consider the multiplicatively-renormalizable massless scalar
electrodynamics  in curved space-time.
The form of the renormalization-group improved scalar effective potential
is derived in this theory, paying special attention to the non-minimal
scalar-gravitational sector. In Section {\bf III} we analyze inflation in
frames of above scalar quantum electrodynamics in Jordan frame. We
explicitly derive the slow-roll parameters, the spectral index and the
tensor-to-scalar ratio showing how the quantum corrections enter in these
expressions. We compute the $e$-folds number and we demonstrate that the
model leads to a viable inflationary scenario according with the last
Planck and BICEP2/Keck Array data.
In Section {\bf IV} we consider multiplicatively-renormalizable
higher-derivative gravity coupled with scalar electrodynamics.
The complicated expression for RG improved effective potential in such
theory (with account of QG corrections) is obtained.
 Section {\bf V} is devoted to the study of
QG-induced inflation in comparison with the simplified case of scalar
electrodynamics analyzed before.
QG does support the realization of inflation. Also in this case, we
carefully investigated how the QG corrections enter in the expressions for
the slow-roll parameters, the spectral index and the tensor-to-scalar
ratio.
It is found that the
bound of the Hubble parameter describing the quasi-de Sitter solution of
inflation is influenced by the correction of the mass scale of the theory.
As a consequence, in order to obtain a realistic scenario, the early-time
acceleration results to be weaker when the mass decreases. Conclusions and
final remarks
are given in Section {\bf VI}.

\section{Effective potential in QUANTUM scalar ELECTRODYNAMICS IN CURVED
SPACE-TIME}

In this section, we present  the  renormalization-group (RG) improved
effective potential for  a  massless scalar electrodynamics in curved
space-time~\cite{El,Odeff}.
The general action for multiplicatively-renormalizable  higher-derivative
gravity  can be written as \cite{B,Fradkin}
\begin{equation}
I=\int_\mathcal{M} d^4 x\sqrt{-g}\left[\frac{R}{2\kappa^2}-\Lambda+a_1
R^2+a_2 C_{\mu\nu\xi\sigma}C^{\mu\nu\xi\sigma}+a_3 G+a_4\Box R+\mathcal
L_m\right]\,,
\label{startaction}
\end{equation}
where $g$ is the determinant of the metric tensor $g_{\mu\nu}$,
$\mathcal{M}$ represents
the space-time manifold, $R$ is the Ricci scalar, $\Lambda$ a (positive)
cosmological constant, $\mathcal L_m$ encodes the matter contributions and
$\Box\equiv g^{\mu\nu}\nabla_{\mu}\nabla_{\nu}$ is the
covariant d'Alembertian, with ${\nabla}_{\mu}$ being the covariant derivative
operator associated with the metric. Moreover, $G$ is the Gauss-Bonnet
four-dimensional topological invariant and
$C_{\mu\nu\xi\sigma}C^{\mu\nu\xi\sigma}$ is the ``square'' of the Weyl
tensor,
\begin{equation}
G=R^2-4R_{\mu\nu}R^{\mu\nu}+R_{\mu\nu\xi\sigma}R^{\mu\nu\sigma\xi}\,,\quad
C_{\mu\nu\xi\sigma}C^{\mu\nu\xi\sigma}=\frac{1}{3}R^2-2R_{\mu\nu}R^{\mu\nu}+R_{\xi\sigma\mu\nu}R^{\xi\sigma\mu\nu}\,,\label{Gauss}
\end{equation}
$R_{\mu\nu}\,,R_{\mu\nu\xi\sigma}$ being the
the Ricci tensor and the Riemann tensor, respectively.

In the above expression, $a_{1,2,3,4}$ are dimensionless parameters, while
$1/\kappa^2$ has the dimension of the square of a mass. At  present epoch
we know that it has to be $1/\kappa^2=M_{Pl}^2/8\pi$, $M_{Pl}$ being the
Planck mass.
As usually we assume the parameters $\kappa^2\,,\Lambda\,,a_{1,2,3,4}$ to
be constant, then
the contribution of the Gauss-Bonnet and of the surface term $\Box R$ drop
down, and the action takes the simplified form,
\begin{equation}
I=\int_\mathcal{M}^4\sqrt{-g}\left[
\frac{R}{2\kappa^2}-\Lambda
+a_1 R^2+a_2C_{\mu\nu\xi\sigma}C^{\mu\nu\xi\sigma}+\mathcal L_m\right]\,.
\label{actionR2}
\end{equation}
At the early-time universe, the  matter Lagrangian contains gauge fields,
scalar multiplets  and spinors and the related interactions typical of any
Grand Unified Theory (GUT).
In what follows,
we consider  massless scalar quantum electrodynamics (QED), whose
Lagrangian in curved space-time reads~\cite{pot1, pot2, pot3},
\begin{equation}
\mathcal L_m=-D_\mu\phi D^{\mu} \phi-\frac{1}{4}F^{\mu\nu}F_{\mu\nu}
+\frac{1}{2}
\xi R\phi^2-\frac{1}{4!}f\phi^4\,.\label{mL}
\end{equation}
Here, $D_\mu=\partial_\mu-e A_\mu$ is the covariant derivative,
$F_{\mu\nu}=\partial_\mu A_\nu-\partial_\nu A_\mu$ is the electromagnetic
tensor, $\xi\,,f$ are dimensionless coupling constants, and $\phi$ is a
complex scalar field. The effective Lagrangian  reads
\begin{equation}
\mathcal
L_m=-\frac{\partial_\mu\phi\partial^{\mu}\phi}{2}-V_\text{eff}(\phi, R)\,,
\end{equation}
where $\phi=\sqrt{|\phi|}$, while the effective potential
$V_\text{eff}\equiv V_\text{eff}(\phi, R)$ has to be evaluated
in one-loop approximation
in the background where $\phi$ and $R$ are almost constants. It satisfies
the standard RG equation,
\begin{equation}
\left[
\mu\frac{\partial}{\partial\mu}+
\beta_{e^2}(t')\frac{\partial}{\partial e^2 (t')}
+\beta_{f}(t')\frac{\partial}{\partial f(t')}
+\beta_{\xi}(t')\frac{\partial}{\partial \xi(t')}
-\gamma(t')\phi(t')\frac{\partial}{\partial\phi(t')}
\right]V_\text{eff}=0\,.\label{const}
\end{equation}
In this expression, couplings $e^2(t')\,,f(t')\,,\xi(t')$ and $\phi(t')$
are the functions of the renormalization parameter
$t'$ given by
\begin{equation}
t'=\frac{1}{2}\log \left[\frac{\phi^2}{\mu^2}\right]\,,
\label{tprime}
\end{equation}
where $\mu$ is a mass parameter in the range
$\mu\sim\mu_{GUT}=10^{15}\text{GeV}$. We point out that $\mu<M_{Pl}\simeq
1.2\times 10^{19}\text{GeV}$, and during inflation $1< \phi^2/\mu^2$.
Moreover,
$\beta_{e^2, f, \xi}(t')$ and $\gamma(t')$ are the corresponding
beta-functions, namely~ (see works on RG-improved effective potential in
flat and curved spacetime \cite{beta1,El})
\begin{eqnarray}
\beta_{e^2}(t')&=&\frac{2e^4(t')}{3(4\pi)^2}\,,\quad
\beta_f(t')=\frac{1}{(4\pi)^2}\left(\frac{10}{3}f(t')^2-12e(t')^2 f(t')+36
e(t')^4\right)\,,\nonumber\\
\beta_\xi(t')&=&\frac{\left(\xi(t')-\frac{1}{6}\right)}{(4\pi)^2}\left(\frac{4}{3}f(t')-6e(t')^2\right)\,,\quad
\gamma(t')=-\frac{3e^2(t')}{(4\pi)^2}\,.\label{betadue}
\end{eqnarray}
One finds that Eq.~(\ref{const}) can be recasted in the form
\begin{equation}
V_\text{eff}\equiv V_\text{eff}(\mu\text{e}^{t'}, e^2(t'), f(t'),\xi(t'),
\phi(t'))\,,
\end{equation}
such that
\begin{equation}
\frac{d e^2(t')}{d t'}=\beta_{e^2}(t')\,,\quad
\frac{d f(t')}{d t'}=\beta_f(t')\,,\quad \frac{d \xi(t')}{d
t'}=\beta_\xi(t')\,,\quad
\frac{d\phi(t')}{d t'}=-\gamma(t')\phi(t')\,.\label{betauno}
\end{equation}
Thus, one derives
\begin{equation*}
e(t')^2=e^2\left(1-\frac{2e^2 t'}{3(4\pi)^2}\right)^{-1}\,,\quad
f(t')=\frac{e(t')^2}{10}\left[
\sqrt{719}\tan\left[
\frac{\sqrt{719}}{2}\log e(t')^2+C
\right]+19
\right]\,,
\end{equation*}
\begin{equation}
\xi(t')=\frac{1}{6}+\left(\xi-\frac{1}{6}\right)\left(\frac{e(t')^2}{e^2}\right)^{-26/5}
\frac{\cos^{2/5}[\sqrt{719}(\log e^2)/2+C]}{\sqrt{719}(\log
e^2(t'))/2+C}\,,\quad
\phi^2(t')=\phi^2\left(1-\frac{2e^2
t'}{3(4\pi)^2}\right)^{-9}\,,\label{valoriuno}
\end{equation}
where we set $e\equiv e(t'=0)$, $f\equiv f(t'=0)$, $\xi\equiv\xi(t'=0)$,
$\phi\equiv\phi(t'=0)$ and
\begin{equation*}
C=\arctan\left[
\frac{1}{\sqrt{719}}\left(\frac{10f}{e^2}-19\right)
-\frac{1}{2}\sqrt{719}\log e^2
\right]\,.
\end{equation*}
Finally, one rewrites the effective potential $V_\text{eff}$ in the form
\begin{equation}
V_\text{eff}=-\frac{1}{4!}f(t')\phi^4(t')+\frac{1}{2}\xi(t')
R\phi^2(t')\,.\label{effectiveLm}
\end{equation}
By plugging the corresponding expressions for the effective coupling
constants, one gets for small $t'$ and weak coupling the following
one-loop effective potential,
\begin{equation}
V_\text{eff}=-\tilde
f\phi^4-A\phi^4\left[\log\frac{\phi^2}{\mu^2}-\frac{25}{6}\right]+\tilde\xi
R\phi^2-B R\phi^2\left[\log\frac{\phi^2}{\mu^2}-3\right]\,,\label{Veff}
\end{equation}
with
\begin{eqnarray}
\tilde f = \frac{f}{4!}\,,\quad
\tilde\xi=\frac{\xi}{2}\,,\quad
A = \frac{1}{48(4\pi)^2}\left(\frac{10}{3}f^2+36e^4\right)\,,\quad
B = \frac{1}{12(4\pi)^2}
\left[
\left(\xi-\frac{1}{6}\right)
\left(\frac{4f}{3}-6e^2\right)+6\xi e^2
\right]\,.\label{set1}
\end{eqnarray}
This result is valid for $\phi$ and therefore $R$ almost constants.
Moreover, $\mu^2$ represents the scale of inflation (we assume that when
$\phi^2=\mu^2$ inflation ends). In the next section, we  use the
Lagrangian (\ref{actionR2}) with $\Lambda=0$ and constant coefficients in
the gravitational sector. Note that we work in Jordan frame through this
paper.

\section{Inflation in scalar quantum electrodynamics}

It is interesting  to see how the model can reproduce the early-time
inflation at the GUT scale. Note that RG-improved effective potential has
been applied for the study of inflation in Refs.~\cite{El,rginfl, OdEl}.
Actually, the inflation due to scalar QED  has been already studied in
Ref.~\cite{OdEl} in the Einstein frame, but here we  work in the Jordan
frame. This is due to the fact that account of quantum corrections breaks
the mathematical equivalence between Einstein and Jordan frames\cite{qel}.
Hence, the inflationary predictions  from QFT like the case under
consideration maybe significally different. Furthermore, generally
speaking there is no even classical equivalence between Jordan and
Einstein frames in the presence of Weyl-squared term.
We also mention that the study of  RG improved inflationary scalar
electrodynamics and $SU(5)$ scenarios confronted with Planck 2013 and
BICEP2 results can be found in Ref.~\cite{OdEl}.

Let us consider the flat Friedmann-Robertson-Walker (FRW) space-time
described by the metric
\begin{equation}
ds^2=-d t^2+a^2(t)d{\bf x}^2\,,
\end{equation}
$a\equiv a(t)$ being the scale factor of the universe. We immediatly note
that the square of the Weyl tensor in (\ref{actionR2}) is identically null
and does not give any contribution to the dynamics of the model.
We will also set the cosmological constant term $\Lambda=0$.
If the field $\phi\equiv\phi(t)$ depends on the cosmological time only,
the equations of motion (EOMs) are derived as
\begin{equation}
\frac{3 H^2}{\kappa^2}+12 a_1 H^2 R =a_1
R^2+\frac{\dot\phi^2}{2}+\left[V_\text{eff}-R\frac{d V_\text{eff}}{d
R}\right]+
6H^2\frac{d V_\text{eff}}{d R}
-3 H\dot F\,,\label{EOM1}
\end{equation}
\begin{equation}
-2 F \dot H=\dot\phi^2+\ddot F -H \dot F\,.\label{EOM2}
\end{equation}
Here, $H=\dot a/a$ is the Hubble parameter, the dot denotes the time
derivative, $V_\text{eff}$ is given by (\ref{Veff})--(\ref{set1}) and we
introduced the following notation,
\begin{equation}
F\equiv F(R,\phi)=\frac{1}{\kappa^2}+4 a_1 R-2\frac{d V_\text{eff}}{d
R}\,.\label{Fprime}
\end{equation}
From (\ref{EOM1})--(\ref{EOM2}) we also infer the continuity equation of
the scalar field,
\begin{equation}
\ddot\phi+3H\dot\phi=-V'_\text{eff}\,,\label{cons}
\end{equation}
with
\begin{equation}
V'_\text{eff}\equiv\frac{d V_\text{eff}}{d\phi}\,.
\end{equation}
Inflation is commonly described by a (quasi) de Sitter solution in
slow-roll approximation regime ($\dot\phi^2\ll V_\text{eff}$,
$0<V_\text{eff}$, and $|\ddot\phi| \ll |H\dot\phi|$), when Eq.(\ref{EOM1})
and Eq.~(\ref{cons}) take the form
\begin{equation}
\frac{3 H^2}{\kappa^2}\simeq
\left[V_\text{eff}-6H^2\frac{d V_\text{eff}}{d R}\right]
\,,\quad
3H\dot\phi\simeq-V'_\text{eff}\,,
\label{sr}
\end{equation}
where  $R\simeq 12H^2$. In the limit $1\ll a_1\kappa^2 R$ one recovers the
chaotic inflation of the Starobinsky-like models~\cite{Staro, mieistaro,
Zergstaro} in the Jordan frame with Eq.~(\ref{EOM1}) asymptotically
satisfied for a given boundary value of the Hubble parameter. Here, we
assume that $a_1 R\kappa^2$ is not asymptotically dominant.
Thus, from the first equation above,
one derives the de Sitter solution,
\begin{equation}
H_\text{dS}^2\simeq
\frac{\left[\tilde
f+A\left[\log\left[\frac{\phi^2}{\mu^2}\right]-\frac{25}{6}\right]\right]\kappa^2\phi^4}
{-3 +
6\left[\tilde\xi-B\left[\log\left[\frac{\phi^2}{\mu^2}\right]-3\right]\right]\kappa^2\phi^2}\,.
\label{dS1}
\end{equation}
We immediatly see that $H_\text{dS}^2$ is large as long as,
\begin{equation}
1 \ll \tilde\xi\kappa^2\phi^2\rightarrow
\frac{M_{Pl}^2}{\tilde\xi}\ll\phi^2\,.\label{condphi}
\end{equation}
In general, since the field exceeds the Planck mass during inflation, we
must also require that
$\tilde f/\tilde\xi<1$.
From the second equation in (\ref{sr}) we obtain
\begin{eqnarray}
\dot\phi\simeq
\frac{2\phi\left[12H^2\left[-2B-\tilde\xi+B\log\left[\frac{\phi^2}{\mu^2}\right]\right]+
\left[-22 A/3+2\tilde
f+2A\log\left[\frac{\phi^2}{\mu^2}\right]\right]\phi^2\right]}{3H}\,.
\end{eqnarray}
This result is valid when the slow-roll approximation
$\dot\phi^2/V_\text{eff}\ll 1$ holds true, namely,
\begin{eqnarray}
\frac{\dot\phi^2}{V_\text{eff}}
\simeq
-\frac{4\left[2(\tilde f-25A/6)-4B(\tilde f-25A/6)\kappa^2\phi^2
-2A (\tilde\xi+3B)
\kappa^2\phi^2-2A\left[-1+B\kappa^2\phi^2\right]\log\left[\frac{\phi^2}{\mu^2}\right]\right]^2}
{3\kappa^2\phi^2\left[\tilde f+A\left[
\log\left[\frac{\phi^2}{\mu^2}\right]-\frac{25}{6}\right]\right]^2\left[-1-2\tilde\xi\kappa^2\phi^2+2B\left[\log\left[\frac{\phi^2}{\mu^2}\right]-3\right]\kappa^2\phi^2\right]}\ll
1\,.\label{dotphiV}
\end{eqnarray}
Since the quantum corrections encoded in $A\,,B$ are small,
\begin{eqnarray}
\frac{\dot\phi^2}{V_\text{eff}}
\sim\frac{16}{3\kappa^2\phi^2+6\tilde\xi\kappa^4\phi^4}\,,
\end{eqnarray}
and
(\ref{dotphiV}) is well satisfied by taking into account (\ref{condphi}).

To study perturbations left at the end of inflation, one needs the
``slow-roll'' parameters~\cite{sr, corea},
\begin{equation}
\epsilon_1=-\frac{\dot
H}{H^2}\,,\quad\epsilon_2=\frac{\ddot\phi}{H\dot\phi}\,,\quad
\epsilon_3=\frac{\dot F}{2 H F}\,,\quad
\epsilon_4=\frac{\dot E}{2 H E}\,,\label{srpar}
\end{equation}
where
\begin{equation}
E=F+\frac{3\dot F^2}{2\dot\phi^2}\,.
\end{equation}
The slow-roll parameters at the first order in $A$ and $B$ are
obtained\footnote{
Note that, by using (\ref{EOM2}), asymptotically one must
find~\cite{miofRphi,miofRphi2},
\begin{equation*}
\epsilon_4=
\frac{\left[\frac{\dot\phi^2}{H\dot F(R,\phi)}\left(-4\epsilon_3\right)
+6\epsilon_1+6\epsilon_3(1-\epsilon_2)
\right]}{2\left[\frac{\dot\phi^2}{H\dot F(R,\phi)}+3\epsilon_3\right]}\,.
\end{equation*}
However, in our model
\begin{equation*}
\frac{H \dot
F(R,\phi)}{\dot\phi^2}\simeq\frac{\kappa^2\phi^2(\tilde\xi^2-4a_1\tilde
f)}{\tilde\xi}
+\frac{50 A a_1\kappa^2\phi^2 }{3\tilde\xi}
+\frac{6B\kappa^2\phi^2}{\tilde\xi}
-\frac{3B\kappa^2\phi^2(\tilde\xi^2-4a_1\tilde f)}{\tilde\xi^2}
\,,
\end{equation*}
diverges as $\sim\kappa^2\phi^2$ like $\epsilon_1\,,\epsilon_3$, rendering
$\epsilon_1\simeq -\epsilon_3$ in the limit $1\ll\tilde\xi\kappa^2\phi^2$
(otherwise, $\epsilon_4\simeq (\epsilon_1/\epsilon_3+1)$ results to be
large) and the expression above for $\epsilon_4$ is useless (it holds true
only at the zero order respect to $\epsilon_{1,2,3})$.
} under the condition (\ref{condphi}),
\begin{eqnarray}
\epsilon_1\simeq
\frac{4}{\kappa^2\phi^2}+\frac{4A(2-\tilde\xi\kappa^2\phi^2)}{\tilde
f\kappa^2\phi^2}
+8B\left(-1+\frac{1}{\tilde\xi\kappa^2\phi^2}\right)\,,
\nonumber
\end{eqnarray}
\begin{eqnarray}
\epsilon_2\simeq\frac{2}{\tilde\xi\kappa^4\phi^4}
+\frac{2A(-3+4\tilde\xi\kappa^2\phi^2)}{\tilde f\kappa^2\phi^2}
+8B\left(2-\frac{1}{\tilde\xi\kappa^2\phi^2}\right)\,,
\nonumber
\end{eqnarray}
\begin{eqnarray}
\epsilon_3\simeq
-\frac{4}{\kappa^2\phi^2}-
\frac{4A(8a_1\tilde f-\tilde\xi(4a_1\tilde f-\tilde
\xi^2)\kappa^2\phi^2)}{\tilde f \kappa^2\phi^2(4a_1\tilde f-\tilde \xi^2)}
+8B\left(1+\frac{(4a_1\tilde
f+\tilde\xi^2)}{\kappa^2\phi^2(\tilde\xi^3-4a_1\tilde
f\tilde\xi)}\right)\,,
\nonumber
\end{eqnarray}
\begin{eqnarray}
\epsilon_4\simeq-\frac{4}{\kappa^2\phi^2}+2A\left(\frac{2\tilde\xi}{\tilde
f}-
\frac{4(240a_1^2 \tilde f^2+4a_1 \tilde
f(1-18\tilde\xi)\tilde\xi+3\tilde\xi^4)}
{\tilde f(48 a_1\tilde f+\tilde\xi-12\tilde\xi^2)(4a_1\tilde
f-\tilde\xi^2)\kappa^2\phi^2}\right)
+8B\left(1+\frac{(4a_1\tilde
f+\tilde\xi^2)}{\kappa^2\phi^2(\tilde\xi^3-4a_1\tilde
f\tilde\xi)}\right)\,.\label{sr1}
\end{eqnarray}
We see that in the first approximation $\epsilon_1\simeq -\epsilon_3$ like
in pure modified gravity.
It is also interesting to note that the $R^2$-term  contributes only in
the one-loop corrections.
This fact is not surprising. The $R^2$-higher derivative term in the
gravitational action may support the de Sitter expansion if it is dominant
(otherwise, like in our case, its contribution disappears from the
Friedmann-like equations with constant Hubble parameter), but does not
drive the exit from inflation (for example, in the Jordan frame of the
Starobinsky model this role is played by the Einstein's term).

The amount of inflation is measured by the $e$-folds number,
\begin{equation}
N:=\log\left[\frac{a(t_\text{f})}{a(t_\text{i})}\right]=\int^{t_\text{f}}_{t_\text{i}}
H dt \,,
\end{equation}
where $t_\text{i, f}$ are the time at the beginning and at the end of
inflation, respectively. In our case we derive
\begin{equation}
N=\int^{\phi_\text{f}}_{\phi_\text{i}}\frac{H}{\dot\phi}d\phi\simeq
\frac{1}{8}\kappa^2\phi_\text{i}^2\,,\label{Nresult}
\end{equation}
where $\phi_\text{i,f}$ are the values of the field at the beginning and
at the end of inflation and we  considered $\kappa^2\phi_\text{e}^2\ll
\kappa^2\phi_\text{i}^2$. In order to obtain the thermalization of
observable universe, it must be $55<N<65$.

The spectral index $n_s$ and the tensor-to-scalar ratio $r$ take into
account the cosmological scalar and tensorial perturbations left at the
end of inflation and
are given by~\cite{corea},
\begin{equation}
n_s=1-4\epsilon_1-2\epsilon_2+2\epsilon_3-2\epsilon_4\,,\quad
r=16(\epsilon_1+\epsilon_3)\,,\label{indexes}
\end{equation}
where $\epsilon_{1,2,3,4}$ must be evaluated in the limit
$\phi=\phi_\text{i}$. Since in our case in first approximation
$\epsilon_1\simeq-\epsilon_3$, we write the whole formula for the
tensor-to-scalar ratio $r$ as,
\begin{equation}
r=-8(3-\sqrt{4n_T+1})\,,\quad
n_T=\frac{(1+\epsilon_3)(2-\epsilon_1+\epsilon_3)}{(1-\epsilon_1)^2}\,,
\label{indexes2}
\end{equation}
which leads to (at the second order in the slow-roll parameters),
\begin{equation}
r\simeq16(\epsilon_1+\epsilon_3)+16\epsilon_1(\epsilon_1+\epsilon_3)\,.
\end{equation}
We get\footnote{In the computation of the tensor-to-scalar ratio we have
taken into account the contribution from $1/(\tilde\kappa^4\phi^4)$ also.}
\begin{eqnarray}
(1-n_s)&\simeq&\frac{16}{\kappa^2\phi^2}
+\frac{4A(192a_1\tilde f+(5-48\tilde\xi)\tilde\xi)}{\tilde f(48 a_1\tilde
f+\tilde\xi-12\tilde\xi^2)\kappa^2\phi^2}+\frac{16B}{\tilde\xi\kappa^2\phi^2}
\,,\nonumber\\
r&\simeq&
\frac{64\tilde\xi}{(4a_1\tilde f-\tilde\xi^2)\kappa^4\phi^4}-
\frac{128 A\tilde\xi^2}{\tilde f(4a_1\tilde f-\tilde\xi^2)\kappa^2\phi^2}
-
\frac{256 B\tilde\xi}{(4a_1\tilde
f-\tilde\xi^2)\kappa^2\phi^2}\,.\label{indexesuno}
\end{eqnarray}
By using the limit $\phi\simeq\phi_\text{i}$ and by plugging  the
$e$-folds number (\ref{Nresult}) one has
\begin{eqnarray}
(1-n_s) &\simeq&\frac{2(1+B/\tilde\xi)}{N}+\frac{A(192a_1\tilde
f+(5-48\tilde\xi)\tilde\xi)}{2\tilde f(48 a_1\tilde
f+\tilde\xi-12\tilde\xi^2)N}\,,\nonumber\\
r&\simeq&
\frac{\tilde\xi}{(4a_1\tilde f-\tilde\xi^2)N^2}-
\frac{16 A\tilde\xi^2}{\tilde f(4a_1\tilde f-\tilde\xi^2)N}
-\frac{32 B\tilde\xi}{(4a_1\tilde f-\tilde\xi^2)N}\,.
\end{eqnarray}
The recent Planck satellite results~\cite{WMAP, Planck} constraint these
quantities as
$n_{\mathrm{s}} = 0.968 \pm 0.006\, (68\%\,\mathrm{CL})$ and
$r < 0.11\, (95\%\,\mathrm{CL})$.
Moreover, the last BICEP2/Keck Array data~\cite{PlanckBICEP2} yield a
(combined) upper limit for the tensor-to-scalar ratio as $r<0.07\,
(95\%\,\mathrm{CL})$.
If one takes
$N\sim 55-65$, in the limit $A=B=0$, the tensor-to-scalar ratio is small
enough to satisfy the Planck and the BICEP2/Keck Array data, while the
spectral index is in agreement with the Planck results inside the given
range. Thus, the one-loop  potential slightly changes these indexes, and
the model is viable as long as $|B/\tilde\xi|\,,|A/\tilde f|\ll 1$.

\section{The one-loop effective potential in quantum scalar electrodynamics
with higher-derivative quantum gravity}

Let us now generalize the results of above section when quantum gravity
(QG) coupled with massless QED is taken into account.
This theory is known to be
multiplicatively renormalizable but the question with its unitarity
remains to be open. In this work we consider such theory as kind of
effective QG model in order to estimate its possible influence to
inflationary universe.
QG corrections to the QED beta-functions   can be found in Ref.~\cite{B},
but the derivation of the effective potential is quite complicated and can
be given only in an implicit form applying linear curvature approximation,
due to the complexity of the one-loop RG equations.

Higher derivative quantum corrections enter in (\ref{betauno}) as
\begin{equation}
\frac{d e^2(t')}{d t'}=\beta_{e^2}(t')\,,\quad
\frac{d f(t')}{d t'}=\beta_f(t')+\Delta\beta_f(t')\,,\quad \frac{d
\xi(t')}{d t'}=\beta_\xi(t')+\Delta\beta_\xi(t')\,,\quad
\frac{d\phi(t')}{d t'}=-\left(\gamma(t')+\Delta_\gamma(t')\right)\phi(t')\,,
\end{equation}
where $\beta_{e^2, f, \xi}(t')$ and $\gamma(t')$ are given by
(\ref{betadue}) and the QG corrections read
\begin{eqnarray}
\Delta\beta_f(t')&=&
\frac{1}{(4\pi)^2}
\left[
\lambda(t')^2\xi(t')^2\left(15+\frac{3}{4\omega(t')^2}
-\frac{9\xi(t')}{\omega(t')^2}+\frac{27\xi(t')^2}{\omega(t')^2}
\right)\right.
\nonumber\\ &&\left.
-\lambda(t') f(t')\left(
5+3\xi(t')^2+\frac{33\xi(t')^2}{2\omega(t')}
-\frac{6\xi(t')}{\omega(t')}+\frac{1}{2\omega(t')}
\right)
\right]\,,
\nonumber\\
\Delta\beta_\xi(t')&=&\frac{1}{(4\pi)^2}\lambda(t')\xi(t')\left[
-\frac{3}{2}\xi(t')^2+4\xi(t')+3+\frac{10}{3}\omega(t')+\frac{1}{\omega(t')}\left(-\frac{9}{4}\xi(t')^2
+5\xi(t')+1\right)
\right]\,,\nonumber\\
\Delta_\gamma(t') &=&
\frac{\lambda(t')}{4(4\pi)^2}
\left[
\frac{13}{3}-8\xi(t')-3\xi(t')^2-\frac{1}{6\omega(t')}-\frac{2\xi(t')}{\omega(t')}+\frac{3\xi(t')^2}{2\omega(t')}
\right]\,.
\end{eqnarray}
Here, $\lambda(t')$ and $\omega(t')$, where only $\lambda(t')$ has an
explicit formulation, correspond to the running coupling constants
$a_1\equiv a_1(t')$ and $a_2\equiv a_2(t')$ in (\ref{actionR2}), which
interact with the matter sector and are given by
\begin{equation}
a_1(t')=-\frac{\omega(t')}{3\lambda(t')}\,,\quad
a_2(t')=\frac{1}{\lambda(t')}\,,
\end{equation}
with
\begin{equation}
\lambda(t')=\frac{\lambda}{1+\frac{203\lambda t'}{15(4\pi)^2}}\,,\quad
\frac{d\omega(t')}{d t'}=\beta_\omega(t')=
-\frac{\lambda(t')}{(4\pi)^2}\left[
\frac{10}{3}\omega(t')^2+\left(5+\frac{203}{15}\right)\omega(t')+\frac{5}{12}+3\left(
\xi(t')-\frac{1}{6}
\right)^2
\right]\,,
\label{omegaeq}
\end{equation}
where $\lambda\equiv \lambda(t'=0)$ and in general $0<\lambda$. The local
gauge invariance prohibites the QG correction to $e^2(t')$, which has the
same form of (\ref{valoriuno}). Now it is possible to find the effective
potential (\ref{effectiveLm}) for higher-derivative QG with scalar QED,
and for small $t'$ and small couplings one derives~\cite{Odeff}
\begin{equation}
V_\text{eff}=-\tilde
f\phi^4-A\phi^4\left[\log\frac{\phi^2}{\mu^2}-\frac{25}{6}\right]+\tilde\xi
R\phi^2-B R\phi^2\left[\log\frac{\phi^2}{\mu^2}-3\right]\,,\label{Veff2}
\end{equation}
with
\begin{eqnarray}
&&\tilde f = \frac{f}{4!}\,,\quad
\tilde\xi=\frac{\xi}{2}\,,\nonumber\\&&
A = \frac{1}{48(4\pi)^2}\left[
\frac{10}{3}f^2+36e^4+\lambda^2\xi^2\left(15+\frac{3}{4\omega^2}-\frac{9\xi}{\omega^2}+\frac{27\xi^2}{\omega^2}\right)
-\lambda
f\left(\frac{28}{3}+18\frac{\xi^2}{\omega}-\frac{8\xi}{\omega}-8\xi+\frac{1}{3\omega}\right)
\right]\,,\nonumber\\&&
B = -\frac{1}{4(4\pi)^2}
\left[
\left(\xi-\frac{1}{6}\right)
\left(\frac{4f}{3}-6e^2\right)+6\xi e^2
+\lambda\xi\left[
8\xi+\frac{5}{6}+\frac{10}{3}\omega+\frac{1}{\omega}\left(-3\xi^2+6\xi+\frac{13}{12}\right)
\right]
\right]\,,\label{conv}
\end{eqnarray}
where, as usually, $\omega\equiv\omega(t'=0)$,
$e\equiv e(t'=0)$, $f\equiv f(t'=0)$, $\xi\equiv\xi(t'=0)$ and
$\phi\equiv\phi(t'=0)$.
In the next section, this expression for the effective potential is
applied to study inflation in higher-derivative QG with scalar QED.

\section{Inflation in quantum gravity with scalar quantum electrodynamics}

In this section, we will analyze the inflation for the effective potential
(\ref{Veff2}) with running coupling constants for the gravitational
Lagrangian in (\ref{actionR2}).
The general formalism of a RG-improved theory requires an explicit
dependence on the renormalization scale of $\kappa^2\equiv\kappa^2(t')$
and $\Lambda\equiv\Lambda(t')$ in (\ref{actionR2}).
In particular, $\kappa^2(t')$ obeys to the differential equation~\cite{B},
\begin{equation}
\frac{d\kappa^2(t')}{d
t'}=\frac{\kappa^2\lambda(t')}{(4\pi)^2}\left[\frac{10\omega(t')}{3}-\frac{13}{6}-\frac{1}{4\omega(t')}\right]\,.\label{eqkappa}
\end{equation}
Despite to the fact that it is not possible to solve explicitly the
equation for $\omega(t')$ in (\ref{omegaeq}), we will try to estimate the
gravitational running coupling constants  by using the fixed points
of this equation, which correspond to\footnote{A derivation of the
adimensional quantity $\kappa(t')^4\Lambda(t')$ can be found in
Ref.~\cite{B}.}
\begin{equation}
\omega_{1,2}=\frac{1}{50}\left[-139\pm\sqrt{2\left(9473+750\tilde\xi-4500\tilde\xi^2\right)}\right]\,,
\label{omega12}
\end{equation}
where  $\xi(t')\simeq\xi$ and we have introduced the notation
in~(\ref{conv}). By perturbing the solution of $\omega(t')$ around the
fixed points as $\omega(t')\simeq\omega_{1,2}+\delta\omega(t')$ with
$|\delta\omega(t')|\ll 1$, from (\ref{omegaeq}) one has,
\begin{equation}
\frac{d\omega(t')}{d t'}\simeq
-\frac{\lambda}{(4\pi)^2\left(1+\frac{203\lambda
t'}{15(4\pi)^2}\right)}\left[
\frac{20}{3}\omega_{1,2}+\left(5+\frac{203}{15}\right)\right]\delta\omega(t')\,,
\end{equation}
whose solution reads
\begin{equation}
\omega(t')\simeq\omega_{1,2}+\frac{c_0}
{\left(1+\frac{203\lambda t'}{15(4\pi)^2}\right)^q}\,,\quad
q=\frac{15}{203}\left[
\frac{20}{3}\omega_{1,2}+\left(5+\frac{203}{15}\right)\right]\,,
\end{equation}
$c_0$ being a constant. The solution does not diverge only if $0<q$ and we
may assume a stable fixed point  for $\omega(t')\simeq \omega_1$ (i.e.,
with the sign plus inside (\ref{omega12})). In this case, from equation
(\ref{eqkappa}) we obtain
\begin{equation}
\kappa^2(t')\simeq\tilde\kappa^2\left(1+\frac{203\lambda
t'}{15(4\pi)^2}\right)^{15 z/203}\,,\quad
z=\left[\frac{10\omega_1}{3}-\frac{13}{6}-\frac{1}{4\omega_1}\right]\,,
\end{equation}
with $\tilde\kappa^2\equiv\kappa^2(t'=0)$.
We must pose $\tilde\kappa^2= 8\pi/M_{Pl}^2$, namely we would like to
recover the Planck mass when quantum effects disappear, and we require
that $0<z$, such that during inflation the mass scale of the theory
decreases.\\
\\
By taking $t'$ small, one can work with the following forms of
$\kappa^2(t')\,, a_1(t')$ inside (\ref{actionR2}),
\begin{equation}
\frac{1}{\kappa^2(t')}=\frac{1}{\tilde\kappa^2}-2m^2 t'\,,\quad
a_1(t')\equiv \tilde a_1+2 b_1 t'\,,
\end{equation}
where $\tilde a_1=a_1(t'=0)$, $b_1$ is an adimensional parameter and
$m^2$ a mass constant such that (during inflation),
\begin{equation}
m^2<\frac{1}{2\tilde\kappa^2 t'}\,.\label{condm}
\end{equation}
Specifically, it is easy to verify that
\begin{equation}
m^2=\frac{1}{2\tilde\kappa^2}\frac{z\lambda}{(4\pi)^2}\,,\quad
\tilde a_1=-\frac{\omega}{3\lambda}\,,\quad
b_1=-\frac{1}{2}\left[\frac{203\omega}{45(4\pi^2)}
\right]\,.
\end{equation}
Finally, by using (\ref{tprime}), we have
\begin{equation}
\frac{1}{\kappa^2}=\frac{1}{\tilde\kappa^2}-m^2\log\left[\frac{\phi^2}{\mu^2}\right]\,,\quad
a_1=\tilde a_1+
b_1\log\left[\frac{\phi^2}{\mu^2}\right]\,.
\end{equation}
As in the previous section, we will set $\Lambda\equiv \Lambda(t')=0$ in
(\ref{actionR2}) and observe that the variation of the square of the Weyl
tensor on FRW metric when $a_2\equiv a_2(t')$ reads
\begin{equation}
\delta I_{C^2}=a_2(t')\delta\left(\sqrt{-g}C^2\right)+
\left(\sqrt{-g}C^2\right)\delta a_2(t')=0\,,
\end{equation}
due to the fact that the square of the Weyl tensor is identically null in
homogeneous and isotropic space-time.
We must note that in the presence of running coupling constants also the
Gauss-Bonnet $G$ and the $\Box R$-terms in the general formulation of the
action (\ref{startaction}) give contribution, but here, for the sake of
simplicity, we will omit such terms.

\begin{figure}[h] \centering
\includegraphics[width=18pc]{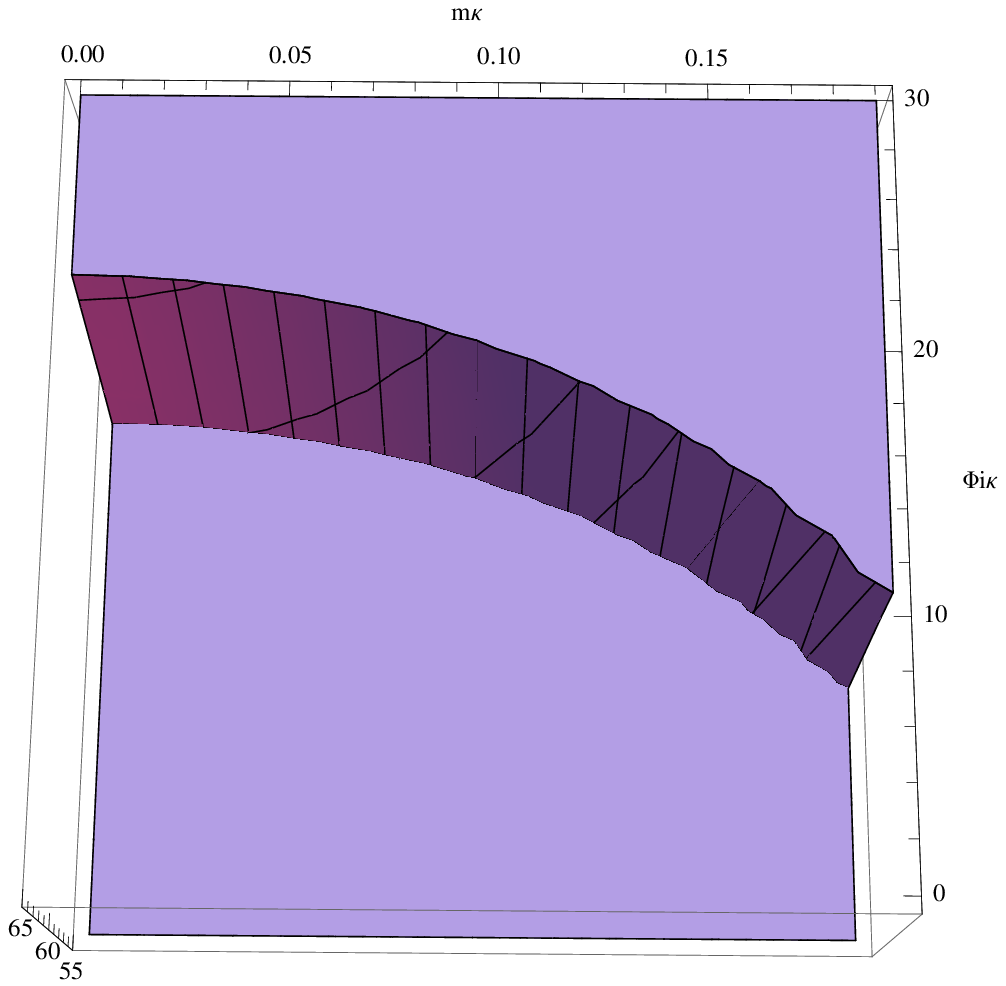}
\includegraphics[width=18pc]{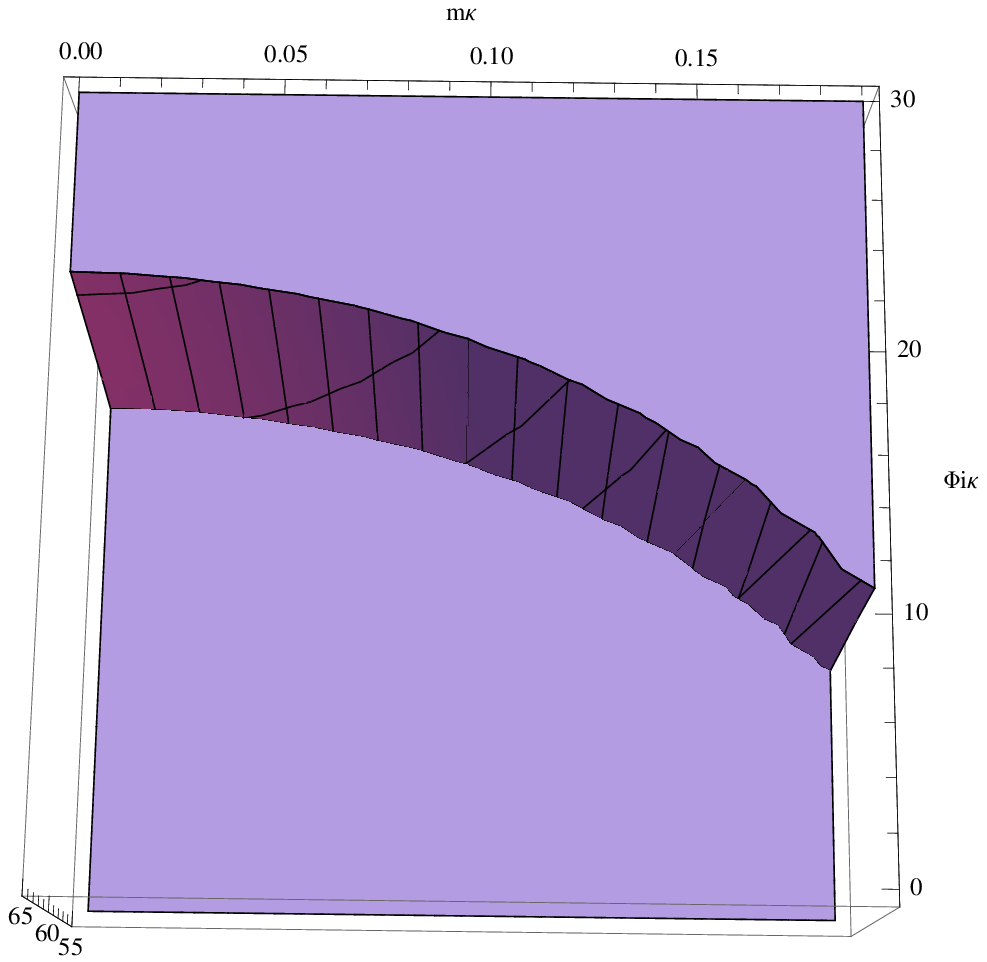}
\caption{The $e$-folds number $N$ (left) and the quantity $2/(1-n_s)$
(right) as functions of $\phi_\text{i}\tilde\kappa$ and $m\tilde\kappa$
for the quantum scalar electrodynamics   with higher-derivative
quantum gravity corrections. The dark zones correspond to $55<N<65$ and
$55<2/(1-n_s)<65$, respectively. We can observe that the values of
$\phi_\text{i}$ and $m$ which lead to a correct amount of inflation ($N$),
also lead to a spectral index consistent with Planck data.}
\label{figuno}
\end{figure}

On flat FRW space-time the first Friedmann equation of the model is
derived as
\begin{eqnarray}
&&3H^2\left[\frac{1}{\tilde\kappa^2}-m^2\log\left[\frac{\phi^2}{\mu^2}\right]\right]
+12  \left[\tilde a_1+b_1\log\left[\frac{\phi^2}{\mu^2}\right]\right] H^2 R
=\nonumber\\
&&\hspace{3cm}
 \left[\tilde a_1+b_1\log\left[\frac{\phi^2}{\mu^2}\right]\right] R^2
\frac{\dot\phi^2}{2}+\left[V_\text{eff}-R\frac{d V_\text{eff}}{d R}\right]+
6H^2\frac{d V_\text{eff}}{d R}
-3 H\dot F\,,\label{EOM1bis}
\end{eqnarray}
with
\begin{equation}
F\equiv
F(R,\phi)=\left[\frac{1}{\tilde\kappa^2}-m^2\log\left[\frac{\phi^2}{\mu^2}\right]\right]
+4 \left[\tilde a_1+b_1\log\left[\frac{\phi^2}{\mu^2}\right]\right]
R-2\frac{d V_\text{eff}}{d R}\,.\label{Fprimebis}
\end{equation}
Moreover, the continuity equation of the scalar  field is given by
\begin{equation}
\ddot\phi+3H\dot\phi=-V'_\text{eff}
+\frac{1}{\phi}\left[-m^2 R+2b_1 R^2\right]\,.\label{consbis}
\end{equation}
In the slow-roll approximation with $R\simeq 12H^2$ the equations
(\ref{EOM1bis}) and (\ref{consbis}) assume the form
\begin{equation}
3
H^2\left[\frac{1}{\tilde\kappa^2}-m^2\log\left[\frac{\phi^2}{\mu^2}\right]\right]\simeq
\left[V_\text{eff}-6H^2\frac{d V_\text{eff}}{d R}\right]
\,,\quad
3H\dot\phi\simeq-V'_\text{eff}+\frac{1}{\phi}\left[-12 H^2 m^2 +288 b_1
H^4\right]\,.
\label{srbis}
\end{equation}
Now the de Sitter solution for the effective potential (\ref{Veff}) is
given by,
\begin{equation}
H_\text{dS}^2\simeq
\frac{\left[\tilde
f+A\left[\log\left[\frac{\phi^2}{\mu^2}\right]-\frac{25}{6}\right]\right]\tilde\kappa^2\phi^4}
{-3\left[1-m^2\tilde\kappa^2\log\left[\frac{\phi^2}{\mu^2}\right]\right] +
6\left[\tilde\xi-B\left[\log\left[\frac{\phi^2}{\mu^2}\right]-3\right]\right]\tilde\kappa^2\phi^2}\,,
\label{dS2}
\end{equation}
and it is large under the condition
\begin{equation}
\frac{\left[1-m^2\tilde\kappa^2\log\left[\frac{\phi^2}{\mu^2}\right]\right]}{\tilde\kappa^2\tilde\xi}\ll\phi^2\,.\label{condphibis}
\end{equation}
If we identify $\tilde\kappa^2=8\pi/M_{Pl}^2$, since the field may be
larger than the Planck mass during inflation, we must also require $\tilde
f/\tilde\xi<1$.
From the second equation in (\ref{srbis}) we get
\begin{eqnarray}
\dot\phi\simeq
\frac{288 b_1 H^4-
12H^2m^2\phi^2+2\phi^2\left[12H^2\left[-2B-\tilde\xi+B\log\left[\frac{\phi^2}{\mu^2}\right]\right]+
\left[-22A/3+2\tilde
f+2A\log\left[\frac{\phi^2}{\mu^2}\right]\right]\phi^2\right]}{3H\phi}\,.
\end{eqnarray}
Thus, by taking $A, B$ and $b_1$ small, we obtain
\begin{eqnarray}
\frac{\dot\phi^2}{V_\text{eff}}
\sim\frac{16\left[1-m^2\tilde\kappa^2\left[\log\left[\frac{\phi^2}{\mu^2}\right]-1\right]\right]^2}{3\tilde\kappa^2\phi^2\left[1+m^2\tilde\kappa^2\log\left[\frac{\phi^2}{\mu^2}\right]\right]+6\tilde\xi\tilde\kappa^4\phi^4}\,,
\end{eqnarray}
which goes to zero when (\ref{condphibis}) is satisfied (the corrections
to $\dot\phi^2/V_\text{eff}$ are at the second order in $b_1$). The
slow-roll parameters, at the first order in $A\,,B$ and $b_1$, read
\begin{eqnarray}
\hspace{-1cm}&&\epsilon_1\simeq
\frac{4\left[1-m^2\tilde\kappa^2\left[\log\left[\frac{\phi^2}{\mu^2}\right]-1\right]\right]}{\tilde\kappa^2\phi^2}+\frac{2A\left[4-m^2\tilde\kappa^2\left[4\log\left[\frac{\phi^2}{\mu^2}\right]-3\right]-2\tilde\xi\tilde\kappa^2\phi^2\right]}{\tilde
f\tilde\kappa^2\phi^2}
+\frac{8B\left[1-m^2\tilde\kappa^2\left[\log\left[\frac{\phi^2}{\mu^2}\right]-1\right]-\tilde\xi\tilde\kappa^2\phi^2\right]}{\tilde
\xi\tilde\kappa^2\phi^2}
\nonumber\\
\hspace{-1cm}&&\hspace{1cm}-\frac{
8b_1\tilde
f\left[-m^2\tilde\kappa^2+2\tilde\xi\tilde\kappa^2\phi^2\right]}{\tilde\xi^2\tilde\kappa^2\phi^2}\,,
\nonumber
\end{eqnarray}
\begin{eqnarray}
\hspace{-1cm}&&\epsilon_2\simeq\frac{8m^2}{\phi^2}
+\frac{2A\left[-3-m^2\tilde\kappa^2\left[1-3\log\left[\frac{\phi^2}{\mu^2}\right]\right]+4\tilde\xi\tilde\kappa^2\phi^2\right]}{\tilde
f\tilde\kappa^2\phi^2}
-\frac{8B\left[1-m^2\tilde\kappa^2\left[\log\left[\frac{\phi^2}{\mu^2}\right]-1\right]-2\tilde\xi\tilde\kappa^2\phi^2\right]}{\tilde
\xi\tilde\kappa^2\phi^2}
\nonumber\\
\hspace{-1cm}&&\hspace{1cm}
+\frac{8b_1\tilde
f\left[-1-m^2\tilde\kappa^2\left[3-\log\left[\frac{\phi^2}{\mu^2}\right]\right]+4\tilde\xi\tilde\kappa^2\phi^2\right]}{\tilde
\xi^2\tilde\kappa^2\phi^2}\,,
\nonumber
\end{eqnarray}
\begin{eqnarray}
\hspace{-1cm}&&\epsilon_3\simeq
-\frac{4\left[1-m^2\tilde\kappa^2\left[\log\left[\frac{\phi^2}{\mu^2}\right]-1\right]\right]}{\tilde\kappa^2\phi^2}
+\frac{2A\left[2\tilde\xi(4a_1\tilde
f-\tilde\xi^2)\tilde\kappa^2\phi^2+4a_1\tilde
f(-4-3m^2\tilde\kappa^2)-m^2\tilde\kappa^2\tilde\xi^2+16a_1\tilde f
m^2\tilde\kappa^2\log\left[\frac{\phi^2}{\mu^2}\right]\right]}{\tilde
f(4a_1\tilde f-\tilde\xi^2)\tilde\kappa^2\phi^2}
\nonumber\\
\hspace{-1cm}&&\hspace{1cm}
+\frac{8B\left[(\tilde\xi^3-4a_1\tilde f\tilde\xi)\tilde\kappa^2\phi^2+
(4a_1\tilde f+\tilde\xi^2)(1-
m^2\tilde\kappa^2\left[\log\left[\frac{\phi^2}{\mu^2}\right]-1\right]
\right]}{(\tilde\xi^3-4a_1\tilde f\tilde\xi)\tilde\kappa^2\phi^2}
\nonumber\\
\hspace{-1cm}&&\hspace{1cm}
\nonumber\\
\hspace{-1cm}&&\hspace{1cm}
-\frac{8b_1\tilde f\left[-2\tilde\xi
(-4a_1\tilde f+\tilde\xi^2)\tilde\kappa^2\phi^2-
4a_1\tilde f
m^2\tilde\kappa^2+\tilde\xi^2(-3-2m^2\tilde\kappa^2)+3m^2\tilde\kappa^2\tilde\xi^2\log\left[\frac{\phi^2}{\mu^2}\right]\right]}{\tilde\xi^2(-4a_1\tilde
f+\tilde\xi^2)\tilde\kappa^2\phi^2}
\,,
\nonumber
\end{eqnarray}
\begin{eqnarray}
\hspace{-1cm}&&\epsilon_4\simeq-\frac{4\left[1-m^2\tilde\kappa^2\left[\log\left[\frac{\phi^2}{\mu^2}\right]-1\right]\right]}{\tilde\kappa^2\phi^2}+2A\left[\frac{2\tilde\xi}{\tilde
f}\right.
\nonumber\\
\hspace{-1cm}&&\hspace{1cm}
-
\frac{
\left[\right.-192a_1^2\tilde f^2(-5-2m^2\tilde\kappa^2)+4a_1\tilde
f\tilde\xi(4-72\tilde\xi-3m^2\tilde\kappa^2(-1+16\tilde\xi))-\tilde\xi^3(-12\tilde\xi-m^2\tilde\kappa^2
(1+24\tilde\xi))}
{\tilde f(192a_1^2\tilde f^2+4a_1\tilde
f(1-24\tilde\xi)\tilde\xi+\tilde\xi^3(-1+12\tilde\xi))\tilde\kappa^2\phi^2}
\nonumber\\
\hspace{-1cm}&&\hspace{1cm}
\left.+
\frac{\left.4m^2\tilde\kappa^2(240a_1^2\tilde f^2+4a_1\tilde
f(1-18\tilde\xi)\tilde\xi
+3\tilde\xi^4)\log\left[\frac{\phi^2}{\mu^2}\right]\right]}
{\tilde f(192a_1^2\tilde f^2+4a_1\tilde
f(1-24\tilde\xi)\tilde\xi+\tilde\xi^3(-1+12\tilde\xi))\tilde\kappa^2\phi^2}
\right]
\nonumber\\
\hspace{-1cm}&&\hspace{1cm}
+8B\left[1-
\frac{(4a_1\tilde f+\tilde\xi^2)\left[
48a_1\tilde
f-\tilde\xi(-1-m^2\tilde\kappa^2+12\tilde\xi)-m^2\tilde\kappa^2(48a_1\tilde
f+\tilde\xi-12\tilde\xi^2)\log\left[\frac{\phi^2}{\mu^2}\right]
\right]}{\tilde\xi
\left(192a_1^2\tilde f^2+4a_1\tilde
f(1-24\tilde\xi)\tilde\xi+\tilde\xi^3(-1+12\tilde\xi)\right)
\kappa^2\phi^2}
\right]
+8b_1\tilde f\left[\frac{2}{\tilde\xi}\right.
\nonumber\\
\hspace{-1cm}&&\hspace{1cm}
+\frac{
\left[
-192a_1^2\tilde f^2(-1-2m^2\tilde\kappa^2)+\tilde\xi^3
(-3-2m^2\tilde\kappa^2+36\tilde\xi)+4a_1\tilde
f\tilde\xi(-48\tilde\xi-m^2\tilde\kappa^2(1+24\tilde\xi))
\right.
}
{\tilde\xi^2(192a_1^2\tilde f^2+4a_1\tilde
f(1-24\tilde\xi)\tilde\xi+\tilde\xi^3(-1+12\tilde\xi))\tilde\kappa^2\phi^2}
\nonumber\\
\hspace{-1cm}&&\hspace{1cm}
\left.
-\frac{\left.3m^2\tilde\kappa^2(64a_1^2\tilde f^2-64a_1\tilde f
\tilde\xi^2+\tilde\xi^3(-1+12\tilde\xi))\log\left[\frac{\phi^2}{\mu^2}\right]
\right]}
{\tilde\xi^2(192a_1^2\tilde f^2+4a_1\tilde
f(1-24\tilde\xi)\tilde\xi+\tilde\xi^3(-1+12\tilde\xi))\tilde\kappa^2\phi^2}
\right]
\,.\label{sr2}
\end{eqnarray}
The $e$-folds is given by
\begin{equation}
N
\simeq
-\int^{\phi_\text{f}}_{\phi_\text{i}}\frac{\tilde\kappa^2\phi}{4-4m^2\tilde\kappa^2\left[\log\left[\frac{\phi^2}{\mu^2}\right]-1\right]}
\simeq
\frac{\kappa^2\phi_\text{i}^2}{8-8m^2\tilde\kappa^2\left[\log\left[\frac{\phi_\text{i}^2}{\mu^2}\right]-1\right]}
\sum_{n=0}^{\tilde n}
\frac{n!(-4m^2\tilde\kappa^2)^n}{\left[4-4m^2\tilde\kappa^2\left[\log\left[\frac{\phi_\text{i}^2}{\mu^2}\right]-1\right]\right]^n}
\,,\label{Nresultbis}
\end{equation}
where we used the fact $\phi_\text{f}\ll \phi_\text{i}$ and we must cut
the series at some $n=\tilde n$. For example, a simple extimation of
$\tilde n$ may be given by
\begin{equation}
\frac{\phi_\text{i}}{\phi_\text{f}}
\simeq
\left[\frac{4-4m^2\tilde\kappa^2\left[\log\left[\frac{\phi_\text{i}^2}{\mu^2}\right]-1\right]}
{4-4m^2\tilde\kappa^2\left[\log\left[\frac{\phi_\text{f}^2}{\mu^2}\right]-1\right]}\right]^{\tilde
n+1}\,,
\end{equation}
namely when we cannot ignore the contributions from $\phi_\text{f}$ due to
the large number of $n$. In the limit $m^2=0$, one recovers
(\ref{Nresult}). We observe that in general, when $\phi_\text{i}$ is large
enough with respect to the mass scale $\mu$, the computation of the
$e$-folds simply leads to
\begin{equation}
N\simeq
\frac{\kappa^2\phi_\text{i}^2}{8-8m^2\tilde\kappa^2\left[\log\left[\frac{\phi_\text{i}^2}{\mu^2}\right]-1\right]}\,.
\end{equation}
By using (\ref{indexes})--(\ref{indexes2}) we are ready to calculate the
spectral index and the tensor-to-scalar ratio of the theory as,
\begin{eqnarray}
&&\hspace{-1.5cm}(1-n_s)\simeq\frac{16\left[1-m^2\tilde\kappa^2\left[\log\left[\frac{\phi^2}{\mu^2}\right]-2\right]\right]}
{\tilde\kappa^2\phi^2}\nonumber\\&&
+\frac{4A\left[-96a_1\tilde
f(-2-3m^2\tilde\kappa^2)+\tilde\xi(5-48\tilde\xi-m^2\tilde\kappa^2(-5+24\tilde\xi))-m^2\tilde\kappa^2
(192a_1\tilde
f+5\tilde\xi-48\tilde\xi^2)\log\left[\frac{\phi^2}{\mu^2}\right]\right]}{\tilde
f(48a_1\tilde f+\tilde\xi-12\tilde\xi^2)\tilde\kappa^2\phi^2}\nonumber\\&&
-\frac{16B\left[48 a_1\tilde f(1+2
m^2\tilde\kappa^2)-\tilde\xi(-1+12\tilde\xi-m^2\tilde\kappa^2)-
m^2\tilde\kappa^2
(48 a_1\tilde
f+\tilde\xi-12\tilde\xi^2)\log\left[\frac{\phi^2}{\mu^2}\right]\right]}{\tilde
\xi(-48a_1\tilde
f-\tilde\xi+12\tilde\xi^2)\tilde\kappa^2\phi^2}\nonumber\\
&&
+\frac{16b_1\tilde f\left[-96 a_1\tilde f
m^2\tilde\kappa^2+\tilde\xi(1-12\tilde\xi-m^2\tilde\kappa^2(-1+36\tilde\xi))-
m^2\tilde\kappa^2
(\tilde\xi-12\tilde\xi^2)\log\left[\frac{\phi^2}{\mu^2}\right]\right]}{\tilde
\xi^2(-48a_1\tilde
f-\tilde\xi+12\tilde\xi^2)\tilde\kappa^2\phi^2}\,,\nonumber\\
&&\hspace{-1.5cm}
r\simeq\frac{64\tilde\xi\left[1-m^2\tilde\kappa^2\left[\log\left[\frac{\phi^2}{\mu^2}\right]-1\right]\right]^2}{(4a_1\tilde
f-\tilde\xi^2)\tilde\kappa^4\phi^4}
-\frac{\left[128A\tilde\xi^2/\tilde f+256B\tilde
\xi\right]\left[1-m^2\tilde\kappa^2\left[\log\left[\frac{\phi^2}{\mu^2}\right]\right]-1\right]}{(4a_1\tilde
f-\tilde\xi^2)\tilde\kappa^2\phi^2}\nonumber\\
&&-\frac{384 b_1\tilde
f\left[1-m^2\tilde\kappa^2\left[\log\left[\frac{\phi^2}{\mu^2}\right]-1\right]\right]}
{(4a_1\tilde f-\tilde\xi^2)\tilde\kappa^2\phi^2}\,.\label{nr2}
\end{eqnarray}
In the limit $m^2=0$, these indexes with (\ref{Nresultbis}) read
\begin{eqnarray}
&&(1-n_s)\simeq
\frac{2(1+B/\tilde\xi)}{N}+\frac{A(192a_1\tilde
f+(5-48\tilde\xi)\tilde\xi)}{2\tilde f(48 a_1\tilde
f+\tilde\xi-12\tilde\xi^2)N}
+\frac{2 b_1\tilde f\left[\tilde\xi(1-12\tilde\xi)\right]}{\tilde
\xi^2(-48a_1\tilde f-\tilde\xi+12\tilde\xi^2)N}\,,\nonumber\\
&&
r\simeq
\frac{\tilde\xi}{(4a_1\tilde f-\tilde\xi^2)N^2}-
\frac{16 A\tilde\xi^2}{\tilde f(4a_1\tilde f-\tilde\xi^2)N}
-\frac{32 B\tilde\xi}{(4a_1\tilde f-\tilde\xi^2)N}
-\frac{19 4 b_1\tilde f}
{4(4a_1\tilde f-\tilde\xi^2)N}\,,\label{nr3}
\end{eqnarray}
and we recover (\ref{indexesuno}) with the contribution of the
$\log$-correction to $R^2$. On the other hand, when $m^2\neq 0$, in the
limit $A=B=0$ one gets
\begin{equation}
(1-n_s)\simeq
\frac{2\left[8-8m^2\tilde\kappa^2\left[-2+\log\left[\frac{\phi^2}{\mu^2}\right]\right]\right]}
{\tilde\kappa^2\phi^2}\,,\quad
r\simeq
\frac{64\tilde\xi\left[1-m^2\tilde\kappa^2\left[\log\left[\frac{\phi^2}{\mu^2}\right]-1\right]\right]^2}{(4a_1\tilde
f-\tilde\xi^2)\tilde\kappa^4\phi^4}\,,
\end{equation}
and in order to satisfy the last Planck satellite results it must be
\begin{equation}
\frac{2}{(1-n_s)}=\frac{\tilde\kappa^2\phi^2}{\left[8-8m^2\tilde\kappa^2\left[\log\left[\frac{\phi^2}{\mu^2}\right]-2\right]\right]}\simeq
60\,.
\end{equation}
In this case, the spectral index lies inside the observed range, while the
tensor-to-scalar ratio is small enough to be in agreement with Planck and
BICEP2/Keck Array data.
When $\phi\simeq\phi_\text{i}$ and $\phi_\text{i}$ is large enough, this
condition is satisfied for $55<N<65$. In Fig.~\ref{figuno} we plot the
$e$-folds number $N$ and the quantity $2/(1-n_s)$ as functions of
$\phi_\text{i}\tilde\kappa$ and $m\tilde\kappa$. The calculation of $N$
has been carried out in numerical way\footnote{Mathematica \textcircled
c.}
by using the integral in (\ref{Nresultbis}). Since at the end of inflation
the quantum gravity corrections disappear, we posed
$\tilde\kappa=\kappa\equiv \sqrt{8\pi}/M_{Pl}$ and
$\mu=10^{-4}/\tilde\kappa$ ($\sim\mu_{GUT}$). The final value of $\phi$
has been set as $\phi_\text{f}=\mu$, while the range of $\phi_\text{f}$
and $m$ have been chosen as $\mu<\phi_\text{f}<10^2/\kappa$ (we remember
that the field can exceed the Planck scale) and $0\leq
m<\left[\tilde\kappa\sqrt{\log[10^4/(\mu^2\tilde\kappa^2)]}\right]^{-1}$
(see condition (\ref{condm})), respectively.
The dark zones in the graphics correspond to $55<N<65$ and
$55<2/(1-n_s)<65$ and confirm  that the values of $\phi_\text{i}$ and $m$
which lead to a correct amount of inflation, also lead to a spectral index
according with Planck results. Thus, in order to have a viable
inflationary scenario, $10/\kappa\leq\phi_\text{f}\leq20/\kappa$ has to
match $0\leq m<0.20/\kappa$. For $m=0$, one obtains $\phi_\text{f}\simeq
20/\kappa$ (in this limit, we have $N\simeq 60$ in (\ref{Nresult})).

We note that in terms of the $e$-folds number, the QG corrections to the
tensor-to-scalar ratio in (\ref{nr2}) assume the same form of
(\ref{nr3}), namely, given $A, B$ and $b_1$ with a correct amount of
inflation, the model leads to the same corrections to the tensor spectral
index.
On the other hand, the much more
involved expression for the spectral index brings it to have a different
form with respect to (\ref{nr3}), namely, by plugging in the spectral
index the expression for the $e$-folds number, it  remains an explicit
dependence on the mass scale $m^2$. In this sense, given $A, B$ and $b_1$
with a correct amount of inflation, the QG effects lead to different
corrections in the spectral index if compare with the case of pure scalar
QED.

The running mass scale of the model influences the bound of the field and
therefore the de Sitter solution of inflation, since from (\ref{dS2}), for
large values of the field, we get
\begin{equation}
H_\text{dS}^2\sim\frac{\tilde f\phi^2}{6\tilde\xi}\,,
\end{equation}
which is the same expression of (\ref{dS1}). Given $\tilde f$ and
$\tilde\xi$, when $m=0$, in order to have $N\simeq60$, the field must be
$\phi\simeq  22/\tilde\kappa^2$, but when $0<m$, to obtain the same amount
of inflation, the field and the Hubble parameter must be smaller. In this
sense, the quantum corrections to the Planck mass bring to a weaker
acceleration during inflation.\\
\\
As in the previous case, the $R^2$-term does not play a significant role
for the exit from inflation. However, an important remark is in order. If
the mass scale of theory essentially decreases at the early-time epoch
due to the quantum corrections, the following condition may be realized
for a subplanckian value of the curvature,
\begin{equation}
1\ll a_1(t')\kappa(t')^2 R\,.
\end{equation}
In this case, equation (\ref{EOM1bis}) is asymptotically satisfied for
some boundary value of the de Sitter Hubble parameter, and one recovers
inflation from $R^2$-gravity with $\log$-corrections (see
Ref.~\cite{Zergstaro}).\\
\\
From the expression of $\epsilon_1$ in (\ref{sr1}) or (\ref{sr2}) we have,
in terms of $N=\log(a_t{\text{f}}/a(t))$,
\begin{equation}
\epsilon_1\simeq\frac{1}{2N}\,.
\end{equation}
By taking into account that $d/dt=-H(N)d/dN$ together with the definition
of $\epsilon_1$, one easily derives the behaviour of the Hubble parameter
during inflation,
\begin{equation}
H(N)^2=H_0^2 N\,,
\end{equation}
where $H_0^2\ll H_\text{dS}^2=H_0^2 N|_{N\simeq 60}$ gives the value of
the Hubble parameter at the end of inflation. Graceful exit occurs when
$N\simeq 0$ and $\epsilon_1$ exceedes the unit.
Thus, the quantum gravity effects will disappear ($\phi\simeq \mu$) and
our gravitational Lagrangian will turn out to be General Relativity plus a
quadratic correction of the Ricci scalar. The behaviour of this model at
the end of inflation has been well investigated in literature, and it has
been demonstrated that it is compatible with the reheating process for
particle production at the beginning of the Fridmann expansion predicted
by General Relativity.

\section{Conclusions}

In this paper, we have analyzed inflation for a quantum scalar
electrodynamics model in curved space-time and for higher-derivative quantum
gravity with scalar electrodynamics.
The RG improved effective potential is calculated for both theories (i.e.
without and with QG corrections) in Jordan frame.
At the FRW universe, the gravitational
action contains  $R^2$-term
beyond the Hilber-Einstein term $R$.
Our analysis has been carried out in the Jordan frame, due to
non-equivalence of quantum corrected Jordan and Einstein frames.

The resulting inflationary scenarios are in agreement with the Planck and
the last
BICEP2/Keck Array data and bring to an amount of inflation compatible with
the thermalization of the observable universe. Note that as it is clearly
seen from the explicit expressions for slow-roll parameters the analysis
of Jordan frame inflation seems to be much more complicated than  the
corresponding analysis in convinient Einstein frame.

When the quadratic $R^2$-term is not asymptotically dominant in the
gravitational action, its contribution appears only via log-corrections in
the spectral index and the tensor-to-scalar ratio, namely it does not play a
significant role in the exit from inflation, like in the Jordan-frame
representation of the Starobinsky-like models.
However, we note that, due to the running mass scale of the theory, the
$R^2$-term may be dominant for a large subplanckian value of the
curvature: in this case we obtain a pure $R^2$-gravitational model with
$\log$-corrections.

Our analysis shows how one-loop QED and QG corrections enter in
the spectral index and in the tensor-to-scalar ratio of the model under
discussion.
The most interesting corrections  in the coupling constants of
the gravitational action from QG effects are related to the running
gravitational constant. Here, we stress that the
viability of the inflationary scenario does not directly require that the
QG correction to $R$ is small, like in the case of the $\log$-quantum
correction to $R^2$ or the one-loop corrections in the effective potential
of the field.
If at the early-time epoch the Planck mass of the theory decreases, the
bound of the field must be smaller to get a realistic inflationary scenario.
As a consequence, also the Hubble parameter of the (quasi) de Sitter
solution describing inflation is smaller leading to a weaker acceleration.
It is interesting to note that it is straitforward to generalize this
study for Standard Model with higher-derivative QG. However, the
corresponding expressions turn out to be much more involved.

\end{document}